\documentclass[aps,showpacs]{revtex4} 
\usepackage{graphicx}

\begin{document}

\title{$r$-modes in Relativistic Superfluid Stars}

\author{Shijun Yoshida}
\email{yoshida@fisica.ist.utl.pt}
\affiliation{Centro Multidisciplinar de Astrof\'{\i}sica -- CENTRA,
           Departamento de F\'{\i}sica, Instituto Superior T\'ecnico,
           Av. Rovisco Pais 1, 1049-001 Lisboa, Portugal}

\author{Umin Lee}
\email{lee@astr.tohoku.ac.jp}
\affiliation{Astronomical Institute, Graduate School of Science,
           Tohoku University, Sendai 980-8578, Japan}

\date{\today}
	
\begin{abstract}
We discuss the modal properties of the $r$-modes of relativistic 
superfluid neutron stars, taking account of the entrainment effects between 
superfluids. In this paper, the neutron stars are assumed to be 
filled with neutron and proton superfluids
and the strength of the entrainment effects between the superfluids are 
represented by a single parameter $\eta$.  
We find that the basic properties of the 
$r$-modes in a relativistic superfluid star are very similar to those 
found for a Newtonian superfluid star. The $r$-modes of 
a relativistic superfluid star are split into two families, ordinary fluid-like 
$r$-modes ($r^o$-mode) and superfluid-like $r$-modes ($r^s$-mode). The two 
superfluids counter-move for the $r^s$-modes, while they co-move for the 
$r^o$-modes. For the $r^o$-modes, the quantity
$\kappa\equiv\sigma/\Omega+m$
is almost independent of the entrainment 
parameter $\eta$, where $m$ and $\sigma$ are the azimuthal
wave number and the oscillation frequency observed by an inertial observer at 
spatial infinity, respectively. For the $r^s$-modes, on the other hand, 
$\kappa$ almost linearly increases with increasing $\eta$. 
It is also found that the radiation driven instability due to the
$r^s$-modes is much weaker than that of the $r^o$-modes because the matter
current associated with the axial parity perturbations almost completely
vanishes.
\end{abstract}


\maketitle

\section{Introduction}

The $r$-mode instability \cite{A98,FM98} belongs to the secular instability attributable to the 
so-called CFS mechanism \cite{C70,JF78,FS78} that
drives unstable various modes of oscillation in a rotating star. 
The $r$-modes excited in neutron stars are expected to work as a mechanism of decelerating
the spin velocity of neutron stars by emitting gravitational waves, which will be
detectable by LIGO 
and other detectors \cite{LOM98,AKS99,OEL98,LB98,AKSt99}. For a recent review, 
see, e.g., \cite{AK01}.

In old and cold neutron stars having the interior temperature 
below the superfluid transition temperatures $T\sim 10^9  {\rm K}$, 
neutrons in the inner crust and neutrons and protons 
in the core are believed to be in superfluid states \cite{ST83}. 
It is therefore likely that most of observed neutron stars, which have cooled down
very quickly via emission of neutrinos \cite{BP79}, have a core 
with superfluids. 
One of the important effects of rotating superfluids in cold neutron stars is mutual friction, 
which provides a strong 
dissipation mechanism for the oscillation modes \cite{mutual}.
The mutual friction is produced by scattering of normal fluid particles off the vortices in 
the rotating superfluids.
Lindblom and Mendell \cite{LM00} calculated the mutual friction damping associated with the $r$-modes
in neutron stars with a superfluid core and
concluded that the mutual friction in the core is 
ineffective to suppress the $r$-mode instability.
Recently, Lee and Yoshida \cite{LY03} have 
studied the $r$-modes in neutron stars with the superfluid core using a 
different numerical method, confirming most of the results obtained by \cite{LM00}. 
Lee and Yoshida \cite{LY03} have also shown that
the $r$-modes are split into ordinary fluid-like $r$-modes and 
superfluid-like $r$-modes, as first suggested by Andersson and Comer 
\cite{AC01}, and that the instability caused by the 
superfluid-like $r$-modes is extremely weak and easily damped by dissipation 
processes in the interior. 
Very recently, Yoshida and 
Lee \cite{YL03} have calculated inertial modes of superfluid neutron stars, and shown that
the inertial modes are also split into two families, namely, ordinary 
fluid-like inertial modes and superfluid-like inertial modes. 
It is quite reasonable since the $r$-modes belong to a subclass of the inertial modes.
Yoshida and Lee \cite{YL03}  
found that all the inertial modes, except for the ordinary fluid-like $r$-modes, are 
strongly damped by the dissipation due to the mutual friction unless the 
entrainment effects between superfluids are extremely weak.

Most of the numerical studies on the $r$-modes and
the inertial modes in superfluid neutron stars have been done 
within the framework of Newtonian dynamics.
Since neutron stars are general
relativistic objects having the typical relativistic factor 
$M(R)/R$ of order of $10^{-1}$ with $M(R)$ and $R$ being the gravitational 
mass and radius of the star, respectively,
it is necessary to extend the Newtonian analyses to
general relativistic ones.
In this paper, employing the two-constituent model developed by Cater and his co-workers 
\cite{C89,CL98,LSC98,CLL99}, and the formalism devised 
by Lockitch, Friedman, and Andersson \cite{LAF01,LFA02} for 
inertial modes of relativistic normal fluid stars,
we derive a general relativistic formulation for inertial mode oscillations in
superfluid neutron stars. 
In Sec.~II we 
present the basic equations employed in this paper for the dynamics of 
relativistic superfluids, and in Sec.~III numerical results are given,
and Sec.~IV is for discussions and conclusions.
In this paper, we employ geometric units, given by $c=G=1$, where $c$ and 
$G$ are the speed of light and the gravitational constant, respectively, and 
sign conventions used in \cite{MTW}.

\section{Formulation}

\subsection{Two-constituent formalism for superfluid dynamics}

We assume neutron stars are composed of 
superfluid neutrons and a mixture of superfluid protons 
and normal fluid electrons.
We assume perfect charge neutrality between 
the protons and the electrons because the plasma frequency of the mixture 
is much higher than the oscillation frequency considered in this 
study \cite{GM91}. The electrons therefore co-move with the protons. 
In other words, we do not need equations for describing the dynamics of the 
electrons. 
In the following, we loosely call the mixture of the protons and
electrons \lq\lq proton\rq\rq.
To describe the dynamics of the superfluids in a neutron star, we employ the 
relativistic two-constituent formalism, which has been
developed by Carter and his co-workers \cite{C89,CL98,LSC98,CLL99}. 
The fundamental quantity of Carter's superfluid 
formalism is the so-called master function. 
Although we have several
choices for the master function, we take, following Comer, 
Langlois, and Lin \cite{CLL99}, the total thermodynamical energy 
density $-\Lambda$ as the master function. The master function $\Lambda$ 
is assumed to depend on the three scalars $n^2=-n_\alpha n^\alpha$, 
$p^2=-p_\alpha p^\alpha$, and $x^2=-p_\alpha n^\alpha$, where $n^\alpha$ 
and $p^\alpha$ are the conserved number density current of the neutrons 
and the protons, respectively.

Once an explicit functional form of the master function is given, the 
energy-momentum tensor $T^\alpha_\beta$ is given by
\begin{equation}
T^\alpha_\beta=\Psi\,\delta^\alpha_\beta+p^\alpha\chi_\beta+
n^\alpha\mu_\beta\,,
\end{equation}
where the scalar $\Psi$ denotes the generalized pressure, defined by 
\begin{equation}
\Psi=\Lambda-n^\alpha\mu_\alpha-p^\alpha\chi_\alpha\,, 
\end{equation}
and the one-forms $\mu_\alpha$ and $\chi_\alpha$ mean the chemical 
potential covectors, defined by
\begin{eqnarray}
\mu_\alpha={\cal B}n_\alpha+{\cal A}p_\alpha\,,\quad   
\chi_\alpha={\cal A}n_\alpha+{\cal C}p_\alpha\,, 
\label{def-coj-mom}
\end{eqnarray}
where 
\begin{eqnarray}
{\cal A}=-{\partial\Lambda\over\partial x^2}\,,\quad 
{\cal B}=-2{\partial\Lambda\over\partial n^2}\,,\quad 
{\cal C}=-2{\partial\Lambda\over\partial p^2}\,.\quad 
\end{eqnarray}
The one-forms $\mu_\alpha$ and $\chi_\alpha$ are conjugate momenta to 
$n^\alpha$ and $p^\alpha$, respectively. 
The quantities  $\mu=(-\mu_\alpha\mu^\alpha)^{1/2}$ 
and $\chi=(-\chi_\alpha\chi^\alpha)^{1/2}$ are interpreted as the chemical potentials 
of the neutrons and the protons, respectively. 
From equation (\ref{def-coj-mom}), 
we can see that the thermodynamical quantity $\cal A$ determines the strength 
of the entrainment effects between the two superfluids. In other words, if we 
have ${\cal A}=0$, there is no entrainment effect.  
The system of the dynamical equations for the two superfluids is composed 
of two continuity equations, given by   
\begin{equation}
\nabla_\alpha n^\alpha = 0\,,\quad\nabla_\alpha p^\alpha=0\,.
\label{continu-eqs}
\end{equation}
and two Euler equations, given by 
\begin{equation}
n^\alpha\nabla_{[\alpha}\mu_{\beta]}=0\,,\quad 
p^\alpha\nabla_{[\alpha}\chi_{\beta]}= 0\,, 
\label{euler-eqs}
\end{equation}
where $\nabla_\alpha$ means the covariant derivative associated with 
the metric tensor, and the square brackets denote antisymmetrised 
averaging. Since $n^\alpha$ and $p^\alpha$ are the conserved currents, 
it is convenient to introduce two time-like unit vectors $u^\alpha$ and $v^\alpha$
defined by 
\begin{equation}
n^\alpha=n u^\alpha\,,\quad p^\alpha=p v^\alpha\,, 
\end{equation}
where $n$ and $p$ denote the number density of the neutrons and the protons, 
respectively.

\subsection{Equilibrium configurations}

Equilibrium states of a slowly rotating star are described by the stationary 
axisymmetric spacetime, given by the line element 
\begin{eqnarray}
ds^2=g_{\alpha\beta}dx^\alpha dx^\beta=-e^{2\nu(r)}dt^2+e^{2\lambda(r)}dr^2+
r^2(d\theta^2+\sin^2\theta d\varphi^2)-2\omega r^2\sin^2\theta dtd\varphi\,. 
\end{eqnarray}
Here, we have taken account of the rotational effects up to the first order 
of the stellar rotation angular frequency. In this study, we 
assume that the neutron and proton superfluids in equilibrium are in the same 
rotational motion with the uniform rotation angular frequency $\Omega$. The  
four-velocity of the two superfluids is therefore given by 
\begin{equation}
u^\alpha=v^\alpha=\gamma\,(t^\alpha+\Omega\varphi^\alpha)
=e^{-\nu(r)}(t^\alpha+\Omega\varphi^\alpha)\,,
\label{eq-vel}
\end{equation}
where $t^\alpha$ and $\varphi^\alpha$ are the time-like and rotational Killing 
vectors of the spacetime. In virtue of equation (\ref{eq-vel}), the conjugate momentum 
one-forms in equilibrium reduce to 
\begin{eqnarray}
\mu_\alpha&=&({\cal B}n+{\cal A}p)\,u_\alpha=\mu(n,p)\,u_\alpha\,,\nonumber\\
\chi_\alpha&=&({\cal A}n+{\cal C}p)\,u_\alpha=\chi(n,p)\,u_\alpha\,, 
\label{def-chemi0}
\end{eqnarray}
and the energy-momentum tensor of an equilibrium star is given by
\begin{eqnarray}
T^\alpha_\beta=-\Lambda\,u^\alpha u_\beta+\Psi\,q^\alpha_\beta
=-\Lambda\,u^\alpha u_\beta+(\Lambda+\mu n+\chi p)\,q^\alpha_\beta\,,
\end{eqnarray}
where $q^\alpha_\beta$ denotes the projection tensor associated with 
the four-velocity $u^\alpha$, defined by 
$q^\alpha_\beta=\delta^\alpha_\beta+u^\alpha u_\beta$. 

Substituting equations (\ref{def-chemi0}) into Euler equations 
(\ref{euler-eqs}), we obtain the hydrostatic equations 
given by 
\begin{eqnarray}
{1\over\mu}{d\mu\over dr}=-{d\nu\over dr}\,,\quad
{1\over\chi}{d\chi\over dr}=-{d\nu\over dr}\,,\quad
\end{eqnarray}
which leads to
\begin{eqnarray}
{\mu\over\chi}=C\,,
\end{eqnarray}
where $C$ is an integral constant. In this study, we take $C=1$, assuming
that the two superfluids are in chemical equilibrium in the
equilibrium states. 
The metric coefficients are determined from the Einstein equations, which 
are written as 
\begin{eqnarray}
{d\nu\over dr}=e^{2\lambda}\left({M\over r^2}+4\pi r\Psi\right)\,,\quad
{dM\over dr}=-4\pi r^2\Lambda\,,
\end{eqnarray}
\begin{eqnarray}
{d^2\varpi\over dr^2}+\left({4\over r}-{d\nu\over dr}-{d\lambda\over dr}\right)
\,{d\varpi\over dr}-{4\over r}\,\left({d\nu\over dr}+{d\lambda\over dr}\right)
\varpi=0\,,
\end{eqnarray}
where 
\begin{eqnarray}
M(r)={r\over 2}(1-e^{-2\lambda})\,,\quad \varpi=\Omega-\omega\,.
\end{eqnarray}

In this paper, we neglect a normal fluid envelope (e.g., \cite{CLL99}), and
we require that the two superfluids have the common outer 
surface whose radius $R$ is defined by equations $\mu(R)=m_n$ 
and $\chi(R)=m_p$, where $m_n$ and $m_p$ stand for the rest mass of the 
neutron and the proton, respectively. The surface boundary conditions for the
equilibrium structure are given as follows: 
\begin{eqnarray}
\nu={1\over 2}\,\ln\left(1-{2M(R)\over R }\right)\,,\quad 
\varpi+{R\over 3}{d\varpi\over dr}=\Omega\,,\quad {\rm at}\quad r=R\,,
\end{eqnarray}
where $M(R)$ stands for the gravitational mass of the star.
The boundary conditions 
at the stellar center are the regularity condition for all the physical 
quantities.

\subsection{Perturbation equations for the two-fluid model}

To describe fluid perturbations in a star, it is convenient to introduce 
two kinds of changes in physical quantities, called the Eulerian 
and the Lagrangian changes \cite{JF78}. An Eulerian change $\delta Q$ is 
the difference between the quantities $Q$ in the perturbed and the unperturbed states at a 
spacetime point. The relation between the
Lagrangian changes and the Eulerian changes are given by
\begin{eqnarray}
\Delta_nQ=\delta Q+L_{\xi_n}Q\,,\quad \Delta_pQ=\delta Q+L_{\xi_p}Q\,,
\label{def-Lag-p}
\end{eqnarray}
where $L_k$ stands for the Lie derivative along 
the vector $k^\alpha$, and we have introduced the two 
distinct Lagrangian displacement vectors $\xi_n^\alpha$ and $\xi_p^\alpha$
to describe the perturbed motion 
of the neutron and proton superfluids because the two superfluids
can move independently. The Lagrangian displacement is regarded as a 
vector that connects fluid elements in the unperturbed state to the 
corresponding elements in the perturbed state.

We suppose that perturbations of any physical quantity can be expressed in terms of the Lagrangian 
displacements $\xi_n^\alpha$ and $\xi_p^\alpha$, and the Eulerian changes 
in the metric 
$h_{\alpha\beta}=\delta g_{\alpha\beta}$. According to similar consideration 
to that in \cite{JF78}, we can write the Lagrangian changes of the fluid 
velocities of the neutron and proton in terms of the Lagrangian change 
in the metric:
\begin{eqnarray}
\Delta_nu^\alpha=-{1\over 2}u^\alpha u^\mu u^\nu\Delta_ng_{\mu\nu}\,,\quad 
\Delta_pv^\alpha=-{1\over 2}u^\alpha u^\mu u^\nu\Delta_pg_{\mu\nu}\,, 
\end{eqnarray}
where 
\begin{eqnarray}
\Delta_ng_{\alpha\beta}=h_{\alpha\beta}+\nabla_\alpha\xi_{n\beta}+
\nabla_\beta\xi_{n\alpha}\,,\quad
\Delta_pg_{\alpha\beta}=h_{\alpha\beta}+\nabla_\alpha\xi_{p\beta}+
\nabla_\beta\xi_{p\alpha}\,.
\end{eqnarray}
Here, we have used the relation $u^\alpha=v^\alpha$ in the equilibrium. 
In this study, we consider no transfusion between the neutrons and the 
protons. The particle numbers of the neutrons and the protons are 
therefore conserved separately. Then, the conservation equations are 
given by 
\begin{eqnarray}
{\Delta_nn\over n}=-{1\over 2}q^{\alpha\beta}\Delta_ng_{\alpha\beta}\,,\quad 
{\Delta_pp\over p}=-{1\over 2}q^{\alpha\beta}\Delta_pg_{\alpha\beta}\,.
\end{eqnarray}
The Eulerian perturbations of the velocities and the number densities can be  
expressed in terms of $\xi_n^\alpha$, 
$\xi_p^\alpha$, and $h_{\alpha\beta}$ as
\begin{eqnarray}
\delta u^\alpha&=&q^\alpha_\beta L_u\xi_n^\beta+
{1\over 2}u^\alpha u^\mu u^\nu h_{\mu\nu}
=\delta\hat u^\alpha+{1\over 2}u^\alpha u^\mu u^\nu h_{\mu\nu}\,,\nonumber\\
\delta v^\alpha&=&q^\alpha_\beta L_u\xi_p^\beta+
{1\over 2}u^\alpha u^\mu u^\nu h_{\mu\nu}
=\delta\hat v^\alpha+{1\over 2}u^\alpha u^\mu u^\nu h_{\mu\nu}\,,
\end{eqnarray}
\begin{eqnarray}
\delta n=-{n\over 2}q^{\alpha\beta}h_{\alpha\beta}-
q^\alpha_\beta\nabla_\alpha(n\xi_n^\beta)\,,\quad 
\delta p=-{p\over 2}q^{\alpha\beta}h_{\alpha\beta}-
q^\alpha_\beta\nabla_\alpha(p\xi_p^\beta)\,.
\label{p-numb-con-eq}
\end{eqnarray}

According to \cite{C89} and assuming $u^\alpha=v^\alpha$ in the equilibrium, we 
can write the Eulerian perturbations of the conjugate momenta $\mu_\alpha$ 
and $\chi_\alpha$ as 
\begin{eqnarray}
\delta\mu_\alpha&=&\mu\left(\delta\hat u_\alpha+h_{\alpha\beta}u^\beta+
{1\over 2}u_\alpha u^\mu u^\nu h_{\mu\nu}\right)+
p{\cal A}(\delta\hat v_\alpha-\delta\hat u_\alpha)+
\left\{({\cal A}-\bar{\cal A})\delta p+
({\cal B}-\bar{\cal B})\delta n\right\} u_\alpha\,,\nonumber\\
\delta\chi_\alpha&=&\chi\left(\delta\hat v_\alpha+h_{\alpha\beta}u^\beta+
{1\over 2}u_\alpha u^\mu u^\nu h_{\mu\nu}\right)+
n{\cal A}(\delta\hat u_\alpha-\delta\hat v_\alpha)+
\left\{({\cal C}-\bar{\cal C})\delta p+
({\cal A}-\bar{\cal A})\delta n\right\} u_\alpha\,,
\label{def-del-conj}
\end{eqnarray}
where 
\begin{eqnarray}
\bar{\cal A}&=&-2np{\partial{\cal B}\over\partial p^2}-
2n^2{\partial{\cal A}\over\partial n^2}-2p^2{\partial{\cal A}\over\partial p^2}-
np{\partial{\cal A}\over\partial x^2}\,,\nonumber\\
\bar{\cal B}&=&-2n^2{\partial{\cal B}\over\partial n^2}-
4np{\partial{\cal A}\over\partial n^2}-p^2{\partial{\cal A}\over\partial x^2}
\,,\nonumber\\
\bar{\cal C}&=&-2p^2{\partial{\cal C}\over\partial p^2}-
4np{\partial{\cal A}\over\partial p^2}-n^2{\partial{\cal A}\over\partial x^2}\,.
\end{eqnarray}
Note that the second terms on the right-hand side of equations  
(\ref{def-del-conj}) represent the non-dissipative drag force between the 
two superfluids, which 
is proportional to both the function ${\cal A}$ and the velocity difference 
$\delta\hat v_\alpha-\delta\hat u_\alpha$.  
From equation (\ref{euler-eqs}), the perturbed Euler equations are given by 
\begin{eqnarray}
q_\alpha^\beta L_u\delta\mu_\beta&-&q_\alpha^\beta\partial_\beta
(u^\rho\delta\mu_\rho)+\mu\delta\hat u^\beta(\partial_\beta u_\alpha-
\partial_\alpha u_\beta)=0\,,\nonumber\\
q_\alpha^\beta L_u\delta\chi_\beta&-&q_\alpha^\beta\partial_\beta
(u^\rho\delta\chi_\rho)+\chi\delta\hat v^\beta(\partial_\beta u_\alpha-
\partial_\alpha u_\beta)=0\,,
\label{p-euler-eq}
\end{eqnarray}
where $\partial_\alpha$ means the partial derivative.

It is convenient to introduce the vorticity equation when
pulsations in a 
rotating star are considered. Making use of the conjugate 
momenta as dynamical variables, the vorticity equations can be obtained as
\begin{eqnarray}
L_ud\mu_{\alpha\beta}=0\,,\quad L_vd\chi_{\alpha\beta}=0\,,
\end{eqnarray}
where $d\mu_{\alpha\beta}$ and $d\chi_{\alpha\beta}$ are the exterior 
differentiation of the one-forms $\mu_\beta$ and $\chi_\beta$, defined by 
\begin{eqnarray}
d\mu_{\alpha\beta}=\partial_\alpha\mu_\beta-\partial_\beta\mu_\alpha\,,\quad 
d\chi_{\alpha\beta}=\partial_\alpha\chi_\beta-\partial_\beta\chi_\alpha\,.
\end{eqnarray}
Then, the Lagrangian variation of these equations can be derived 
straightforwardly:
\begin{eqnarray}
\Delta_n L_ud\mu_{\alpha\beta}=L_u\Delta_n d\mu_{\alpha\beta}=0\,,\quad 
\Delta_p L_vd\chi_{\alpha\beta}=L_v\Delta_p d\chi_{\alpha\beta}=0\,,
\label{p-vort-eq}
\end{eqnarray}
where $\Delta_n d\mu_{\alpha\beta}$ and $\Delta_p d\chi_{\alpha\beta}$ can be 
written in terms of the Lagrangian changes in the conjugate momenta as 
\begin{eqnarray}
\Delta_n d\mu_{\alpha\beta}=d\Delta_n\mu_{\alpha\beta}=
\partial_\alpha\Delta_n\mu_\beta-\partial_\beta\Delta_n\mu_\alpha\,,\quad
\Delta_p d\chi_{\alpha\beta}=d\Delta_p\chi_{\alpha\beta}=
\partial_\alpha\Delta_p\chi_\beta-\partial_\beta\Delta_p\chi_\alpha\,.
\label{vor-eq}
\end{eqnarray}
Note that since the equations $u^\beta d\Delta_n\mu_{\alpha\beta}=0$ and 
$\partial_{[\alpha}d\Delta_n\mu_{\beta\gamma]}=0$ 
($v^\beta d\Delta_p\chi_{\alpha\beta}=0$ and 
$\partial_{[\alpha}d\Delta_p\chi_{\beta\gamma]}=0$) are satisfied, 
$d\Delta_n\mu_{\alpha\beta}$ ($d\Delta_p\chi_{\alpha\beta}$) has only two 
independent components.

The metric perturbations $h_{\alpha\beta}$ are determined by the linearized 
Einstein equations, given by
\begin{eqnarray}
\delta G^\alpha_\beta=8\pi\delta T^\alpha_\beta\,,
\label{p-ein-eq}
\end{eqnarray}
where $\delta G^\alpha_\beta$ denotes the linearized Einstein tensor. Here,  
$\delta T^\alpha_\beta$ means the linearized energy-momentum tensor, and is 
given by 
\begin{eqnarray}
\delta T^\alpha_\beta=\delta\Psi\delta^\alpha_\beta+\delta n^\alpha \mu_\beta+
\delta p^\alpha \chi_\beta+n^\alpha\delta\mu_\beta+p^\alpha\delta\chi_\beta\,,
\end{eqnarray}
where
\begin{eqnarray}
\delta\Psi={1\over 2}(n^\alpha\mu^\beta+p^\alpha\chi^\beta)h_{\alpha\beta}-
n^\alpha\delta\mu_\alpha-p^\alpha\delta\chi_\alpha\,.
\end{eqnarray}

\subsection{Equations of state: Expanded master function}

In this study, we make use of the same master function as that introduced 
in \cite{ACL02}, which is generally given by 
\begin{equation}
\Lambda(n^2,p^2,x^2)=
\sum_{i = 0}^{\infty}\lambda_i(n^2,p^2)(x^2-np)^i \,.
\label{def-master}
\end{equation}
This expansion of the master function may be justified
because $x^2-np=0$ in equilibrium and the deviation from 
$x^2-np=0$ for perturbed states is the same order of the perturbations.  
Although Comer and Joynt \cite{CJ02} have recently discussed in greater detail
the entrainment effects in general relativistic superfluids, 
we use the master function given above for simplicity. 
In terms of the master function (\ref{def-master}), we can obtain the thermodynamical 
functions in pulsation equations as follows:  
\begin{eqnarray}
{\cal A}&=&-\lambda_1(n^2,p^2)\,,\quad 
{\cal B}=-{p\over n}{\cal A}-
{1\over n}{\partial\lambda_0\over\partial n}\,,\quad
{\cal C}=-{n\over p}{\cal A}-
{1\over p}{\partial\lambda_0\over\partial p}\,,\cr
\bar{\cal A}&=&{\cal A}+{\partial^2\lambda_0\over\partial n\partial p}\,,\quad 
\bar{\cal B}={\cal B}+{\partial^2\lambda_0\over\partial n^2}\,,\quad 
\bar{\cal C}={\cal C}+{\partial^2\lambda_0\over\partial p^2}\,, 
\label{def-abc}
\end{eqnarray}
where we have used the relation $x^2=np$ in the equilibrium. 
Note that $\bar{\cal A}$ becomes equal to ${\cal A}$ if all the expansion 
coefficients $\lambda_i$ are separable in $n$ and $p$ in the sense that
$\lambda_i=f_i(n)+g_i(p)$ for appropriate functions $f_i(n)$ and $g_i(p)$. 
Using the chemical potentials in the equilibrium 
given by
\begin{eqnarray}
\mu =-{\partial\lambda_0\over\partial n}\,,\quad 
\chi=-{\partial\lambda_0\over\partial p}\,,
\end{eqnarray}
we have
\begin{eqnarray}
{\cal A}-\bar{\cal A}={\partial\mu\over\partial p}=
{\partial\chi\over\partial n}\,,\quad
{\cal B}-\bar{\cal B}={\partial\mu\over\partial n}\,,\quad
{\cal C}-\bar{\cal C}={\partial\chi\over\partial p}\,,
\end{eqnarray}
and, assuming all the expansion coefficients 
$\lambda_i$ are separable in $n$ and $p$, we obtain
\begin{eqnarray}
{\cal A}=\bar{\cal A}\,,\quad 
{\cal B}-\bar{\cal B}={d\mu \over dn}\,,\quad
{\cal C}-\bar{\cal C}={d\chi\over dp}\,.
\end{eqnarray}

\subsection{Oscillation equations for inertial modes}

If we assume the star in the equilibrium is stationary and axisymmetric,
the time and azimuthal dependence of the perturbations can be given by
$\exp(i\sigma t+im\varphi)$, where $\sigma$ is 
the oscillation frequency observed by an inertial observer at spatial infinity, 
and $m$ is the azimuthal wave number. Because of the 
rotation effects, it is generally impossible to achieve the separation of 
variables for the perturbed quantities.
We expand the perturbations in terms of the tensor 
spherical harmonics with different $l$'s for a given $m$. In this paper, we consider the 
oscillation modes associated with $m\ge 2$ because we are 
interested in the modes that are unstable against gravitational radiation 
reactions. Thus, we can select the so-called Regge-Wheeler gauge in order 
to fix the gauge freedom for the metric perturbations \cite{RW}. Then,  
the metric perturbations are expanded as 
\begin{eqnarray}
h_{\mu\nu}&=&\sum_{l\ge |m|}\left(\matrix{
e^{2\nu}H_{0l}(r)&H_{1l}(r)&0&0 \cr
{\rm sym}&e^{2\lambda}H_{2l}(r)&0&0 \cr 
{\rm sym}&{\rm sym}&K_l(r)r^2&0\cr 
{\rm sym}&{\rm sym}&{\rm sym}&r^2\sin^2\theta K_l(r)}
 \right)Y^m_l(\theta,\varphi)e^{i\sigma t}\,,\nonumber \\
&+&\sum_{l'\ge |m|}\left(\matrix{
0&0&
ih_{0l'}(r){-1\over\sin\theta}\partial_\varphi Y^m_{l'}(\theta,\varphi)&
ih_{0l'}(r)\sin\theta\partial_\theta Y^m_{l'}(\theta,\varphi)\cr
{\rm sym}&0&
ih_{1l'}(r){-1\over\sin\theta}\partial_\varphi Y^m_{l'}(\theta,\varphi)&
ih_{1l'}(r)\sin\theta\partial_\theta Y^m_{l'}(\theta,\varphi) \cr
{\rm sym}&{\rm sym}&0&0\cr
{\rm sym}&{\rm sym}&{\rm sym}&0}
 \right)e^{i\sigma t}\,,
\end{eqnarray}
where $Y^m_l(\theta,\varphi)$ is the spherical harmonic function, and 
$l=|m|+2(k-1)$ and $l'=l+1$ for even modes, and $l=|m|+2k-1$ and $l'=l-1$ 
for odd modes, where $k=1,2,3,\cdots$. The even and odd modes 
are characterized by symmetry and antisymmetry of the 
eigenfunctions with respect to the equatorial plane. The Lagrangian 
displacements $\xi_{n\alpha}$ and $\xi_{p\alpha}$ can be expanded in terms of 
the vector spherical harmonics as 
\begin{eqnarray}
i\kappa\Omega \xi_{nt}=0\,,\quad i\kappa\Omega \xi_{pt}=0\,,
\end{eqnarray}
\begin{equation}
i\kappa\Omega \xi_{nr}=\sum_{l\ge |m|}{e^{2\lambda}\over r}W_{nl}(r)
Y^m_l(\theta,\varphi)e^{i\sigma t}\,,\quad 
i\kappa\Omega \xi_{pr}=\sum_{l\ge |m|}{e^{2\lambda}\over r}W_{pl}(r)
Y^m_l(\theta,\varphi)e^{i\sigma t}\,,
\label{def-w}
\end{equation}
\begin{eqnarray}
i\kappa\Omega \xi_{n\theta}&=&\sum_{l,l'\ge |m|}
\left(V_{nl}(r)\partial_\theta Y^m_l(\theta,\varphi)-
iU_{nl'}{1\over\sin\theta}\partial_\varphi Y^m_{l'}(\theta,\varphi)
\right)e^{i\sigma t}\,,\nonumber\\
i\kappa\Omega \xi_{p\theta}&=&\sum_{l,l'\ge |m|}
\left(V_{pl}(r)\partial_\theta Y^m_l(\theta,\varphi)-
iU_{pl'}{1\over\sin\theta}\partial_\varphi Y^m_{l'}(\theta,\varphi)
\right)e^{i\sigma t}\,,
\label{def-vu1}
\end{eqnarray}
\begin{eqnarray}
i\kappa\Omega \xi_{n\varphi}&=&\sum_{l,l'\ge |m|}
\left(V_{nl}(r)\partial_\varphi Y^m_l(\theta,\varphi)+
iU_{nl'}\sin\theta\partial_\theta Y^m_{l'}(\theta,\varphi)
\right)e^{i\sigma t}\,,\nonumber\\
i\kappa\Omega \xi_{p\varphi}&=&\sum_{l,l'\ge |m|}
\left(V_{pl}(r)\partial_\varphi Y^m_l(\theta,\varphi)+
iU_{pl'}\sin\theta\partial_\theta Y^m_{l'}(\theta,\varphi)
\right)e^{i\sigma t}\,,
\label{def-vu2}
\end{eqnarray}
where $\kappa\Omega\equiv\sigma+m\Omega$ is the oscillation frequency 
in the corotating frame of the star. 
Note that the gauge freedom for the Lagrangian 
displacement is fixed so as to satisfy $\xi_{nt}=\xi_{pt}=0$. The Eulerian 
perturbations of the number densities for the neutrons and protons are given by 
\begin{eqnarray}
\delta n=\sum_{l\ge |m|}\delta n_l(r)Y^m_l(\theta,\varphi)e^{i\sigma t}\,,\quad
\delta p=\sum_{l\ge |m|}\delta p_l(r)Y^m_l(\theta,\varphi)e^{i\sigma t}\,.
\end{eqnarray}

When we are interested in inertial modes in a slowly rotating 
superfluid star in the lowest order in the rotation frequency $\Omega$,
it is rather easy to generalize the formulation devised by Lockitch et al. 
\cite{LAF01} for relativistic inertial modes in normal fluid neutron stars.
According to \cite{LAF01}, we assume that 
the perturbations satisfy the following ordering laws in the slow rotation limit of 
$\Omega\rightarrow0$: 
\begin{eqnarray}
H_1&=&O(1)\,,\quad W_n=O(1)\,,\quad W_p=O(1)\,,\quad V_n=O(1)\,,\quad 
V_p=O(1)\,,\cr 
H_0&=&O(\Omega)\,,\quad H_2=O(\Omega)\,,\quad K=O(\Omega)\,,\quad 
\delta n=O(\Omega)\,,\quad \delta p=O(\Omega)\,,
\label{g-mode}
\end{eqnarray}
\begin{eqnarray}
h_0=O(1)\,,\quad U_p=O(1)\,,\quad U_n=O(1)\,,
\label{triv-mode}
\end{eqnarray}
and $\kappa=O(1)$. 
Comer \cite{GC02} showed that solutions obeying these ordering laws
are allowed in the perturbation equations for rotating stars, 
exploring the so-called zero-frequency 
subspace of eigensolutions to perturbation equations in a non-rotating 
relativistic superfluid star. If we consider no rotational effects, the 
solutions subject to the ordering laws (\ref{g-mode}) are interpreted as 
infinitely degenerate $g$-modes of zero-frequency, while the solutions 
subject to the ordering laws (\ref{triv-mode}) are consider to be a relativistic 
and superfluid counterpart of the trivial toroidal mode in a non-rotating 
Newtonian normal fluid star \cite{Uetal}.

Substituting the perturbed quantities into linearized equations
(\ref{p-numb-con-eq}), (\ref{p-vort-eq}), and (\ref{p-ein-eq}) and assuming 
the ordering laws for the eigenfunctions (\ref{g-mode}) and 
(\ref{triv-mode}), we can obtain a system of infinitely coupled ordinary 
differential equations for inertial modes in a slowly rotating superfluid star.
To write down the oscillation equations for the inertial modes, 
it is convenient to use vector notation for the 
eigenfunctions ${\bf w}_n$, ${\bf w}_p$, ${\bf v}_n$, ${\bf v}_p$, 
${\bf u}_n$, ${\bf u}_p$, $\bf h$, and $\bf H$, whose components are given by 
\begin{eqnarray}
w_{n,k}=W_{nl}\,,\quad w_{p,k}=W_{pl}\,,\quad v_{n,k}=V_{nl}\,,\quad 
v_{p,k}=V_{pl}\,,\quad u_{n,k}=U_{nl'}\,,\quad u_{p,k}=U_{pl'}\,,\quad
h_k=h_{0l'}\,,\quad H_k=H_{1l}\,,
\end{eqnarray}
where $l=|m|+2(k-1)$, $l^\prime=l+1$ for even modes and $l=|m|+2k-1$,
$l^\prime=l-1$ for odd modes, where $k=1,~2,~3, \cdots$. 
The particle number conservation equations (\ref{p-numb-con-eq}) 
for the neutron and proton are then written as
\begin{eqnarray}
e^{2\nu}r{d(e^{-2\nu}{\bf w}_n)\over dr}&=& -\left(1+{r\over n}{dn\over dr}+
r{d\lambda\over dr}+2r{d\nu\over dr}\right){\bf w}_n+
\Lambda_0 {\bf v}_n\,, \nonumber \\
e^{2\nu}r{d(e^{-2\nu}{\bf w}_p)\over dr}&=& -\left(1+{r\over p}{dp\over dr}+
r{d\lambda\over dr}+2r{d\nu\over dr}\right){\bf w}_p+ 
\Lambda_0 {\bf v}_p\,. 
\label{dif-w}
\end{eqnarray}
Independent components of the vorticity equations (\ref{p-vort-eq}) 
lead to   
\begin{eqnarray}
&&L_1{\bf u}_n-qM_0{\bf v}_n+\bar{q}K_0{\bf w}_n+
{\cal A}_1({\bf u}_p-{\bf u}_n)+{\bf h}=0\,,\nonumber \\
&&L_1{\bf u}_p-qM_0{\bf v}_p+\bar{q}K_0{\bf w}_p+
{\cal A}_2({\bf u}_n-{\bf u}_p)+{\bf h}=0\,,
\label{alg-eq1}
\end{eqnarray}
\begin{eqnarray}
L_0e^{2\nu}r{d(e^{-2\nu}{\bf v}_n)\over dr}&-&
qM_1e^{2\nu}r{d(e^{-2\nu}{\bf u}_n)\over dr}-
m\bar{q}\Lambda_0^{-1}e^{2\nu}r{d(e^{-2\nu}{\bf w}_n)\over dr}-
\left(e^{2\lambda}+mr{d\bar{q}\over dr}\Lambda_0^{-1}\right){\bf w}_n
\nonumber\\
&+&\left(\bar{q}K_1\Lambda_1-r{dq\over dr}M_1\right){\bf u}_n+
m\left(\bar{q}-r{dq\over dr}\Lambda_0^{-1}\right){\bf v}_n-r{\bf H}\nonumber\\
&+&r{d{\cal A}_1\over dr}({\bf v}_p-{\bf v}_n)+{\cal A}_1
\left(e^{2\lambda}({\bf w}_n-{\bf w}_p)+e^{2\nu}
r{d\over dr}(e^{-2\nu}{\bf v}_p-e^{-2\nu}{\bf v}_n)\right)=0\,, \nonumber \\
L_0e^{2\nu}r{d(e^{-2\nu}{\bf v}_p)\over dr}&-&
qM_1e^{2\nu}r{d(e^{-2\nu}{\bf u}_p)\over dr}-
m\bar{q}\Lambda_0^{-1}e^{2\nu}r{d(e^{-2\nu}{\bf w}_p)\over dr}-
\left(e^{2\lambda}+mr{d\bar{q}\over dr}\Lambda_0^{-1}\right){\bf w}_p
\nonumber\\
&+&\left(\bar{q}K_1\Lambda_1-r{dq\over dr}M_1\right){\bf u}_p+
m\left(\bar{q}-r{dq\over dr}\Lambda_0^{-1}\right){\bf v}_p-r{\bf H}\nonumber\\
&+&r{d{\cal A}_2\over dr}({\bf v}_n-{\bf v}_p)+{\cal A}_2
\left(e^{2\lambda}({\bf w}_p-{\bf w}_n)+e^{2\nu}
r{d\over dr}(e^{-2\nu}{\bf v}_n-e^{-2\nu}{\bf v}_p)\right)=0\,,
\label{dif-u}
\end{eqnarray}
where
\begin{eqnarray}
q=2\kappa^{-1}{\varpi\over\Omega}\,,\quad 
\bar{q}=\kappa^{-1}{e^{2\nu}\over r}{d\over dr}
\left(e^{-2\nu}r^2{\varpi\over\Omega}\right)\,,\quad 
{\cal A}_1={p\over\mu}{\cal A}\,,\quad {\cal A}_2={n\over\chi}{\cal A}\,.
\end{eqnarray}
The linearized Einstein equations are reduced to 
\begin{eqnarray}
r{\bf H}=-\Lambda_0^{-1}16\pi e^{2\lambda}r^2(n\mu{\bf w}_n+p\chi{\bf w}_p)\,, 
\label{alg-eq2}
\end{eqnarray}
\begin{eqnarray}
r{d\over dr}\left(r{d\over dr}{\bf h}\right)-\left(1+r{d\nu\over dr}+
r{d\lambda\over dr}\right)r{d\over dr}{\bf h}-\left(e^{2\lambda}\Lambda_1+
2\left(1-e^{2\lambda}+r{d\nu\over dr}+r{d\lambda\over dr}\right){\bf 1}
\right){\bf h}=16\pi e^{2\lambda}r^2(n\mu{\bf u}_n+p\chi{\bf u}_p)\,,
\label{dif-h}
\end{eqnarray}
where {\bf 1} stands for the unit matrix. Here, the matrices $L_0$, $L_1$, 
$K_0$, $K_1$, $\Lambda_0$, $\Lambda_1$, $M_0$, and $M_1$ are defined 
in Appendix. 
The oscillation equations we use are given as a set of linear ordinary 
differential equations for the variables ${\bf w}_n$, ${\bf w}_p$, ${\bf u}_n$,
 ${\bf u}_p$, $\bf h$, and $r{d{\bf h}\over dr}$, which are obtained by 
eliminating the vectors ${\bf v}_n$, ${\bf v}_p$, and $\bf H$ in equations 
(\ref{dif-w}), (\ref{dif-u}), and (\ref{dif-h}) using the algebraic equations 
(\ref{alg-eq1}) and (\ref{alg-eq2}).

The boundary conditions imposed at the center of the star are regularity
condition that all the perturbation functions are regular and do not diverge at $r=0$.
This implies that the eigenfunctions we solve must vanish at $r=0$.
In order to determine the boundary conditions at the stellar surface,
on the other hand, we follow the arguments similar to those given in \cite{CLL99}.
In our specific equilibrium models, for which we have assumed an equation of state 
that is separable in $n$ and $p$,
the generalized pressure $\Psi$ of the fluids can be written as
$\Psi(n,p)=\Psi_n(n)+\Psi_p(p)$, where $\Psi_n(n)$ and $\Psi_p(p)$ represent
the pressures due to the neutron and proton superfluids, respectively. Since the
two superfluids can flow independently, as appropriate surface boundary
conditions we require that the Lagrangian changes
in each of the pressures vanish at the free surface of the star, which leads
to 
\begin{eqnarray}
\Delta_n \Psi_n={d\Psi_n\over dn}\,\Delta_nn=0\,,\quad
\Delta_p \Psi_p={d\Psi_p\over dp}\,\Delta_np=0\,.
\end{eqnarray}
The conditions $\Delta_nn=0$ and $\Delta_pp=0$ at the surface therefore
result in the surface boundary conditions given by $W_{nl}=0$ and $W_{pl}=0$
for inertial mode solutions in the lowest order of $\Omega$.
In the exterior of the star, the only non-trivial metric perturbations
$h_{0l'}$ are determined by equation (\ref{dif-h}), which has two independent
solutions, one is regular at spatial infinity, the other singular.
The regular solution can be given as a series expansion \cite{LAF01,LFA02}:
\begin{eqnarray}
h_{0l'}=\sum_{s=0}^\infty \hat h_{l',s}\left({R\over r}\right)^{l'+s}\,,
\label{exp-sol-h}
\end{eqnarray}
where
\begin{eqnarray}
\hat h_{l',s}={(l'+s-2)!(l'+s+1)!(2l'+1)!\over s!(l'-2)!(l'+1)!(2l'+s+1)!}\,
\left({2 M(R)\over R}\right)^s \hat h_{l',0}\,,
\end{eqnarray}
and $\hat h_{l',0}$ is an arbitrary constant.
If we require that the interior solution
$h_{0l'}$ must be continuous with the exterior solution
(\ref{exp-sol-h}) at the stellar surface because the spacetime must be
regular everywhere, we obtain the boundary conditions for the metric
perturbations
at the stellar surface given by
\begin{eqnarray}
\lim_{\epsilon\rightarrow 0}\left[h_{0l'}(R+\epsilon)\,r{d\over dr}
h_{0l'}(R-\epsilon)-r{d\over dr} h_{0l'}(R+\epsilon)\,
h_{0l'}(R-\epsilon)\right] =0\,.
\end{eqnarray}
For numerical computation, oscillation equations of a finite dimension are 
obtained by disregarding the terms with $l$ larger than 
$l_{max}$ in the expansions of the perturbations, where $l_{max}$ 
is determined so that the eigenfrequency and the eigenfunctions are well 
converged as $l_{max}$ increases. We solve the oscillation equations of a 
finite dimension as an eigenvalue problem with the scaled oscillation 
frequency $\kappa$ using a Henyey type relaxation method (see, e.g., 
\cite{Uetal}).

\section{Numerical results} \label{numres}

\subsection{Equilibrium models}

For the equilibrium structure with $x^2-np=0$, we use a generalized polytropic equation of state,
introduced by Comer et al. \cite{CLL99}, which is given by  
\begin{eqnarray}
\lambda_0(n^2,p^2)=-m_nn-\sigma_nn^{\beta_n}-m_pp-\sigma_pp^{\beta_p}\,,
\label{lam0}
\end{eqnarray}
where $\sigma_n$, $\sigma_p$, $\beta_n$, and $\beta_p$ are constants.
Note that this master function is separable in $n$ and $p$.
The generalized pressure and chemical potentials in the equilibrium can then be 
written as
\begin{eqnarray}
\Psi&=&\sigma_n(\beta_n-1)n^{\beta_n}+\sigma_p(\beta_p-1)p^{\beta_p}\,,\cr
\mu &=&m_n+\sigma_n\beta_nn^{\beta_n-1}\,,\cr
\chi&=&m_p+\sigma_p\beta_pp^{\beta_p-1}\,.
\label{chemi-p}
\end{eqnarray}
Because chemical equilibrium is assumed in the equilibrium state, 
we can analytically write $p$ in terms of $n$ as
\begin{eqnarray}
p=\left({\sigma_n\beta_n\over \sigma_p\beta_p}\right)^{1/(\beta_p-1)}
n^{(\beta_n-1)/(\beta_p-1)}\,,
\label{n-p-rel}
\end{eqnarray}
where we have assumed $m_p=m_n$.
We confirm from equations (\ref{chemi-p}) and (\ref{n-p-rel}) that two 
superfluids in an equilibrium state given by equation (\ref{lam0}) have the 
common outer surface if $\beta_n$ and $\beta_p$ satisfy the conditions 
$\beta_n\ge 1$ and $\beta_p\ge 1$.
In order to introduce the entrainment effects in the perturbed states with $x^2-np\not=0$, 
we include, according to \cite{ACL02}, the expansion coefficient 
$\lambda_1$, given by
\begin{eqnarray}
\lambda_1=-\eta{m_nm_p\over m_pp+\eta(m_nn+m_pp)}\,,
\end{eqnarray}
where $\eta$ is a parameter whose appropriate range is considered to be  
$0.04\le\eta\le0.2$ \cite{S76,BJK96,ACL02}. In this paper, we call $\eta$ the
entrainment parameter.
Making use of the formulas obtained in the preceding sections, we can explicitly 
write the relevant thermodynamical coefficients as follows: 
\begin{eqnarray}
{\cal A}=\bar{\cal A}=\eta{m_nm_p\over m_pp+\eta(m_nn+m_pp)}\,.
\end{eqnarray}

In order to avoid unnecessary singular behaviors of $p$ and $n$ at the stellar 
surface, we consider only the case of $\beta_n=\beta_p$. 
Thus, all the models we use are non-stratified models in the sense that 
$n/p={\rm constant}$ inside the star. The constant parameters
$\sigma_n$, $\sigma_p$, $\beta_n$, and $\beta_p$ in $\lambda_0$ for the
equilibrium equation of state (see Table 1) are the same as those used 
for model 1 calculated in \cite{CLL99}. 
In this paper, 
we consider two equilibrium models, whose physical parameters are tabulated in Table I. 
The model I is almost Newtonian whose relativistic factor $M(R)/R$ is 
$M(R)/R= 10^{-3}$, while the model II is a relativistic 
one with the relativistic factor $M(R)/R=0.15$, the value of which is  
similar to those for the models used in \cite{LFA02}. Although the model I is not appropriate 
as a physical model of neutron stars, we use it to compare with 
completely Newtonian calculations \cite{LY03,YL03}.

\subsection{$r$-mode oscillations}

We calculated the $r$-modes of the two equilibrium models for a wide range 
of the entrainment parameter. It is important to note that the $l'=m$ 
fundamental $r$-modes, whose eigenfunction $U_{nm}$ and/or $U_{pm}$ is 
dominating and has no nodes in radial 
direction, are the only $r$-modes we find in this study. This situation is 
quite similar to that for the $r$-modes in a barotropic ordinary fluid star, in 
which neither the overtone $r$-modes with $l^\prime=m$ nor the $r$-modes with $l'\ne m$ are found 
\cite{LF99,YL00a,YL00b,LFA02}. It is generally found that the
$r$-modes in a relativistic superfluid star are split into two families, 
which we call ordinary fluid $r$-modes ($r^o$-modes) and superfluid $r$-modes
($r^s$-modes), according to the notation of Lee and Yoshida \cite{LY03}. 
Because of doubling of the dynamical degrees of freedom for the system of two superfluids,
the mode splitting of this kind also appears in other 
oscillation modes essentially associated with fluid motions
\cite{E88,LM94,L95,AC01,ACL02,PR02,LY03,YL03}. In Figures 1 and 2, we plot 
the scaled frequencies $\kappa\equiv\sigma/\Omega+m$ of the 
$r$-modes associated with $m=2$ and $3$ as a function of the entrainment 
parameter $\eta$. As shown by the figures, although $\kappa$ 
of the $r^s$-modes increases almost linearly with 
increasing $\eta$, $\kappa$ of the $r^o$-modes is almost independent of $\eta$. 
Comparing the two frequency curves for the models I and II, it is found that 
both the $r^o$-mode and $r^s$-mode frequencies are strongly dependent on the 
relativistic factor $M(R)/R$ of the equilibrium models. This is because the 
frequency of the $r$-modes is roughly proportional to an $average$ of the 
function $\varpi$ within the star, which reflects the general relativistic effects such
as the rotational frame dragging and the gravitational redshift.
Similar dependence of $\kappa$ on the parameter $M(R)/R$ for the $r$-modes 
has been found for relativistic ordinary fluid stars 
\cite{YK98,LAF01,SY01,RK02,LFA02,YL02}.

From Table II, where $\kappa$'s for the $r^o$- and $r^s$-modes
associated with $m=2$ and $3$ are tabulated for several values of $\eta$,
we can confirm that $\kappa$'s for the
$r^o$-modes for model I are nearly equal to $2m/l^\prime(l^\prime+1)$, the value
found for the $r$-modes in Newtonian ordinary 
fluid stars. It is important to note that $\kappa$'s for 
the $r^o$- and $r^s$-modes at $\eta=0$ are not equal to each other, and
the difference between the $\kappa$'s increases as 
the relativistic factor $M(R)/R$ is increased.  
In a Newtonian star, however, $\kappa$'s for the $r^o$- and 
$r^s$-modes at $\eta=0$ are equal to each other in the lowest order of $\Omega$, 
as shown in \cite{LY03}.
The reason  
for the difference between the Newtonian and relativistic $r$-modes at $\eta=0$
is that even at $\eta=0$ there still exist couplings between two superfluid motions 
through the Einstein equations in the relativistic case, but no such couplings exist in the 
Newtonian case because all the Newtonian gravitational perturbations vanish in the 
lowest order of $\Omega$.

In order to illustrate how the eigenfunctions of the $r^o$- and $r^s$-modes 
in a relativistic superfluid star behave, it is convenient to introduce a new set of
variables defined by 
\begin{eqnarray}
W_{+l}&=&{n\,W_{nl}+p\,W_{pl}\over n+p}\,,\quad W_{-l}=W_{nl}-W_{pl}\,,
\nonumber\\
V_{+l}&=&{n\,V_{nl}+p\,V_{pl}\over n+p}\,,\quad V_{-l}=V_{nl}-V_{pl}\,,
\\
U_{+l'}&=&{n\,U_{nl'}+p\,U_{pl'}\over n+p}\,,\quad U_{-l'}=U_{nl'}-U_{pl'}\,. 
\nonumber
\end{eqnarray}
In Figures 3 and 4, we display four dominant coefficients 
$U_{+2}$, $W_{+3}$, $U_{+4}$, and $h_{02}$ for the $r^o$-mode, and $U_{-2}$, 
$W_{-3}$, $U_{-4}$, and $h_{02}$ for the $r^s$-mode for the case of $m=2$ and $\eta=0.1$
for model II. 
Here, the 
normalization condition has been given by $U_{nm}=1$ at $r=R$. The coefficients 
that are not displayed in these figures have very small and negligible amplitudes. 
Figures 3 and 4 represent that the basic properties of the eigenfunctions are 
quite similar to those of the $r^o$- and $r^s$-modes in a Newtonian star with 
superfluidity \cite{LY03,YL03}. The neutrons and protons co-move for the
$r^o$-modes, while they counter-move for the $r^s$-modes. It is also found 
in Figure 4 that the metric perturbation $h_{02}$ almost completely vanishes 
for the $r^s$-modes, which is consistent with the fact that $U_{+m}\sim 0$. 
This means that gravitational radiations due to $r^s$-mode oscillations 
are negligible at least in the lowest order in $\Omega$. It is important to note that the
other coefficients $W_{\pm3}$ and $U_{\pm4}$ are not necessarily negligible compared with 
$U_{\pm 2}$ because 
of the general relativistic effects \cite{LAF01,YL02,LFA02}.

\section{Discussions and Conclusions}

In this paper we have studied the modal properties of the $r$-modes of 
relativistic superfluid neutron stars, taking account of the entrainment 
effects between the neutron and proton superfluids. 
To describe the general relativistic dynamics of the superfluids, we employed
the two-constituent formalism developed by Carter and his co-workers 
\cite{C89,CL98,LSC98,CLL99}. We derived the 
perturbation equations for the relativistic inertial modes in neutron stars filled
with the superfluids by generalizing the 
formalism employed by Lockitch, Andersson, and Friedman \cite{LAF01,LFA02} 
for the relativistic inertial modes in normal fluid neutron stars.
We found that the basic properties of the $r$-modes in 
a relativistic star with the two superfluids are very similar to those in a 
Newtonian superfluid star \cite{LY03,YL03}. We confirmed that 
the $r$-modes of relativistic superfluid stars are split into two 
families, ordinary fluid-like $r$-modes ($r^o$-mode) and superfluid-like $r$-modes 
($r^s$-mode). The two superfluids counter-move for the $r^s$-modes, while 
they co-move for the $r^o$-modes. The dimensionless frequency $\kappa$ for 
the $r^o$-modes is almost independent of the entrainment parameter $\eta$. 
For the $r^s$-modes, on the other hand, $\kappa$ almost linearly increases 
with $\eta$. The gravitational radiation driven instability due to the $r^s$-modes 
is much weaker than that of the $r^o$-modes because the matter current associated 
with the axial parity perturbations vanish almost completely for the former.

A solid crust near the surface of the neutron star has a significant influence on
the modal properties of the $r$-modes in the superfluid core, since
the solid crust supports its own oscillation modes \cite{MVH88,LS96,YL02b,MPS01}, 
and resonance phenomena are expected between the $r$-modes in the superfluid core
and the torsional sound waves in the solid crust even in the general relativistic 
context \cite{YL01,LU01}.
Dissipation in the viscous boundary layer at the interface 
between the fluid core and the solid crust is another important issue for the 
$r$-mode instability, as shown by \cite{ekman}. 
How the dissipation in the core-crust interface can be important for the $r$-mode instability will be 
a quite interesting problem when superfluid neutrons and protons 
in the core and superfluid neutrons in the inner crust
are taken into account simultaneously (see, e.g., \cite{LCHE93}). 

Newtonian superfluid neutron stars can support 
infinite number of inertial modes, which are also split 
into two families, ordinary fluid-like inertial modes ($i^o$-mode) and superfluid-like
inertial modes ($i^s$-mode) \cite{LY03,YL03}.
Non-linear couplings between the $r$-modes and
inertial modes will be important to limit the amplitude growth of the $r$-mode instability,
as recently shown by Arras et al. 
\cite{AEL} for Newtonian normal fluid stars.
To investigate the relativistic inertial modes in
superfluid neutron stars will be one of our future studies.

\acknowledgments

SY acknowledges financial support from Funda\c c\~ao para a Ci\^encia e 
Tecnologia (FCT) through project SAPIENS 36280/99.

\section*{Appendix: Matrices used in pulsation equations}

The components of the matrices, $L_0$, $L_1$, $K_0$, $K_1$, $\Lambda_0$, 
$\Lambda_1$, $M_0$, and $M_1$ are given as follows:

\noindent
For even modes,
\[
({L_0})_{i,i} = 1-{mq\over l(l+1)} \, , \quad
({L_1})_{i,i} = 1-{mq\over (l+1)(l+2)} \, ,
\]
\[
({K_0})_{i,i} = \frac{J^m_{l+1}}{l+1} \, , \quad
({K_0})_{i,i+1} = - \frac{J^m_{l+2}}{l+2} \, ,
\]
\[
({K_1})_{i,i} = - \frac{J^m_{l+1}}{l+1} \, , \quad
({K_1})_{i+1,i} = \frac{J^m_{l+2}}{l+2} \, ,
\]
\[
({\Lambda}_0)_{i,i} = l(l+1) \, , \quad
({\Lambda}_1)_{i,i} = (l+1)(l+2) \, ,
\]
\[
({M}_0)_{i,i} = \frac{l}{l+1} \, J^m_{l+1} \, , \quad
({M}_0)_{i,i+1} = \frac{l+3}{l+2} \, J^m_{l+2} \, ,
\]
\[
({M}_1)_{i,i} = \frac{l+2}{l+1} \, J^m_{l+1} \, , \quad
({M}_1)_{i+1,i} = \frac{l+1}{l+2} \, J^m_{l+2} \, ,
\]
where $l=\vert m \vert + 2 i -2 $ for $i = 1,2,3,\cdots $.

\noindent
For odd modes,
\[
({L_0})_{i,i} = 1-{mq\over l(l+1)} \, , \quad
({L_1})_{i,i} = 1-{mq\over l(l-1)} \, ,
\]
\[
({K_0})_{i,i} = - \frac{J^m_l}{l} \, , \quad
({K_0})_{i+1,i} = \frac{J^m_{l+1}}{l+1} \, ,
\]
\[
({K_1})_{i,i} = \frac{J^m_l}{l} \, , \quad
({K_1})_{i,i+1} = - \frac{J^m_{l+1}}{l+1}\, ,
\]
\[
({\Lambda}_0)_{i,i} = l(l+1) \, , \quad
({\Lambda}_1)_{i,i} = l(l-1) \, ,
\]
\[
({M}_0)_{i,i} = \frac{l+1}{l} \, J^m_l \, , \quad
({M}_0)_{i+1,i} = \frac{l}{l+1} \, J^m_{l+1} \, ,
\]
\[
({M}_1)_{i,i} = \frac{l-1}{l} \, J^m_l \, , \quad
({M}_1)_{i,i+1} = \frac{l+2}{l+1} \, J^m_{l+1} \, ,
\]
where $l=\vert m \vert + 2 i -1 $ for $i = 1,2,3,\cdots $. Here, $J^m_l$ is
the function of $m$ and $l$, defined by $J^m_l=[(l^2-m^2)/(4l^2-1)]^{1/2}$, 
and $q=2\varpi/\kappa\Omega$.

\newpage


\begin{table}
\caption{Parameters describing stellar models I and II.}
\begin{tabular}{|c|c|c|}
\hline
 & model I & model II \\
\hline
$\sigma_n/m_n$ & 0.2 & 0.2  \\
$\sigma_p/m_n$ & 0.5 & 0.5  \\
$\beta_n$ & 2.0 & 2.0  \\
$\beta_p$ & 2.0 & 2.0  \\
$n_c$ (fm$^{-3}$) & 0.0025 & 0.672  \\
$p_c$ (fm$^{-3}$) & 0.001 & 0.269  \\
$M/M_\odot$ & 0.009 & 1.033   \\
$R\ (\hbox{km})$ & 13.39 &  10.17  \\
$M(R)/R$ & 0.001 &  0.150  \\
\hline
\end{tabular}
\label{modtab}
\end{table}

\begin{table}
\caption{Scaled frequencies $\kappa$ for the $r^o$- and $r^s$-modes
associated with $m=2$ and $3$.}
\begin{tabular}{|c|c|cccc|}
\hline
$\eta$&model&$r^o$($m=2$)&$r^s$($m=2$)&$r^o$($m=3$)&$r^s$($m=3$)\\
\hline
 0    & I  &0.6663&0.6660&0.4996&0.4995 \\
      & II &0.6094&0.5459&0.4411&0.4194 \\
\hline
 0.04 & I  &0.6663&0.7592&0.4996&0.5694 \\
      & II &0.6094&0.6178&0.4411&0.4754 \\
\hline
 0.1  & I  &0.6663&0.8990&0.4996&0.6743 \\
      & II &0.6094&0.7240&0.4411&0.5585 \\
\hline
 0.2  & I  &0.6663& 1.132&0.4996&0.8491 \\
      & II &0.6094&0.8964&0.4411&0.6942 \\
\hline
\end{tabular} \label{freqtab2}
\end{table}



\begin{figure}[h]
\centering
\includegraphics[height=7cm,clip]{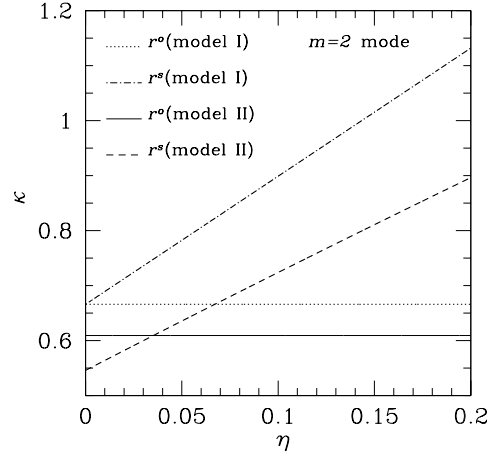}
\caption{$\kappa$'s for $r^o$- and $r^s$-modes associated with $m=2$
in the two models I and II, given as a function of $\eta$.}
\end{figure}

\begin{figure}[h]
\centering
\includegraphics[height=7cm,clip]{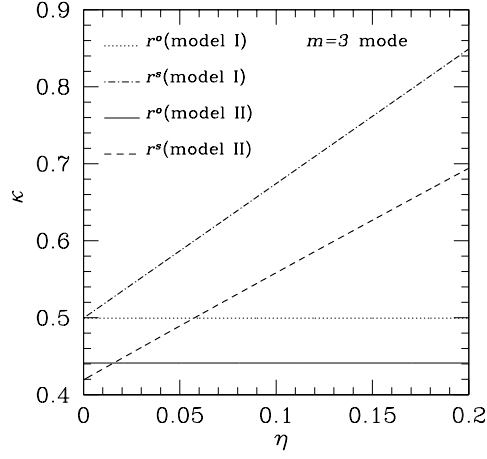}
\caption{Same as Figure 1 but for modes associated with $m=3$.}
\end{figure}

\begin{figure}[h]
\centering
\includegraphics[height=7cm,clip]{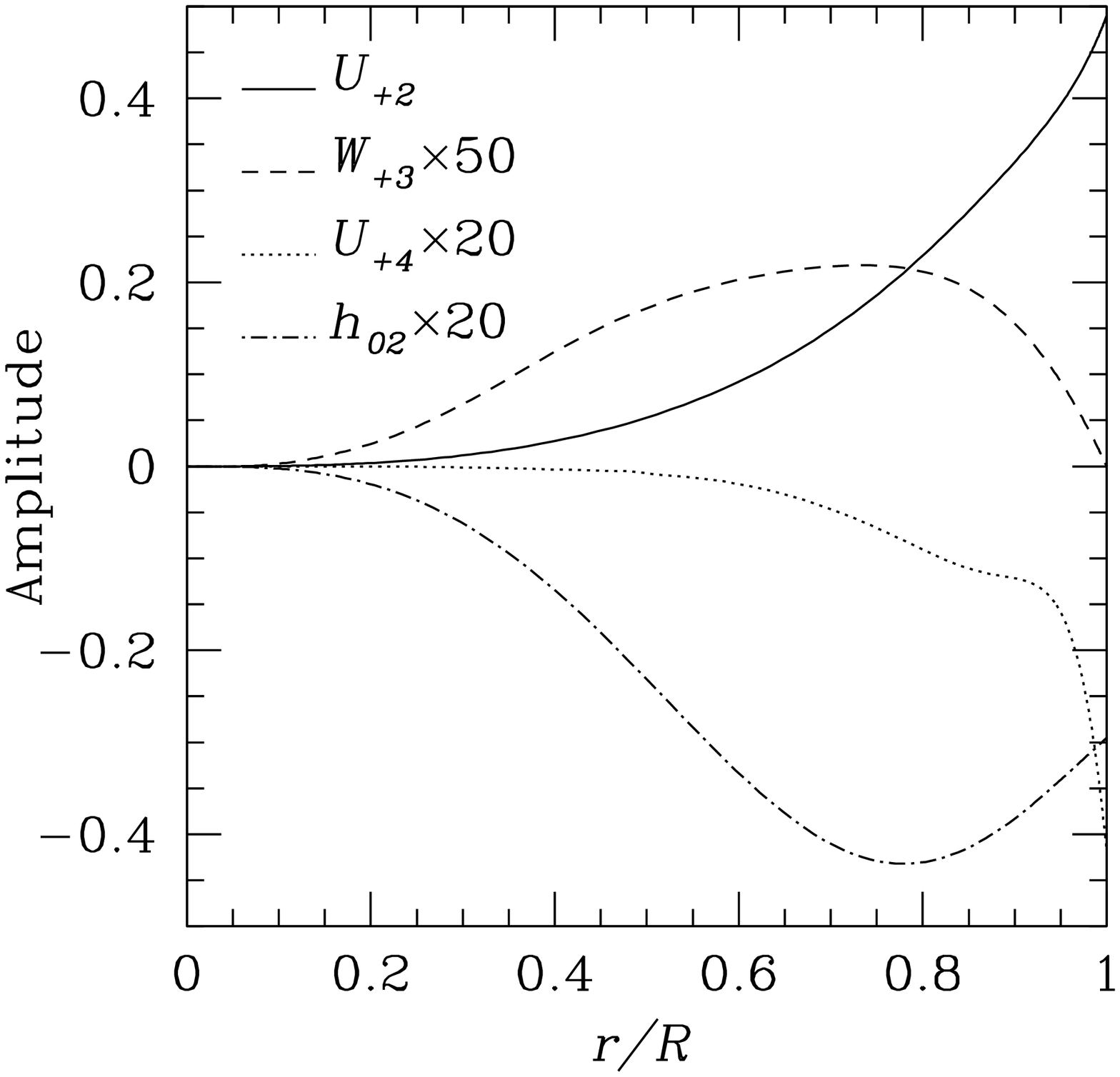}
\caption{Four dominant expansion coefficients $U_{+2}$, $W_{+3}$, $U_{+4}$,
and $h_{02}$ for the $r^o$-mode associated with $m=2$ in model II for
$\eta=0.1$, given as a function of $r/R$. Other coefficients have negligible
amplitude. The corresponding $\kappa$ is given by $\kappa=0.6094$. The 
amplitudes are normalized by $U_{n2}=1$ at $r=R$.}
\end{figure}

\begin{figure}[h]
\centering
\includegraphics[height=7cm,clip]{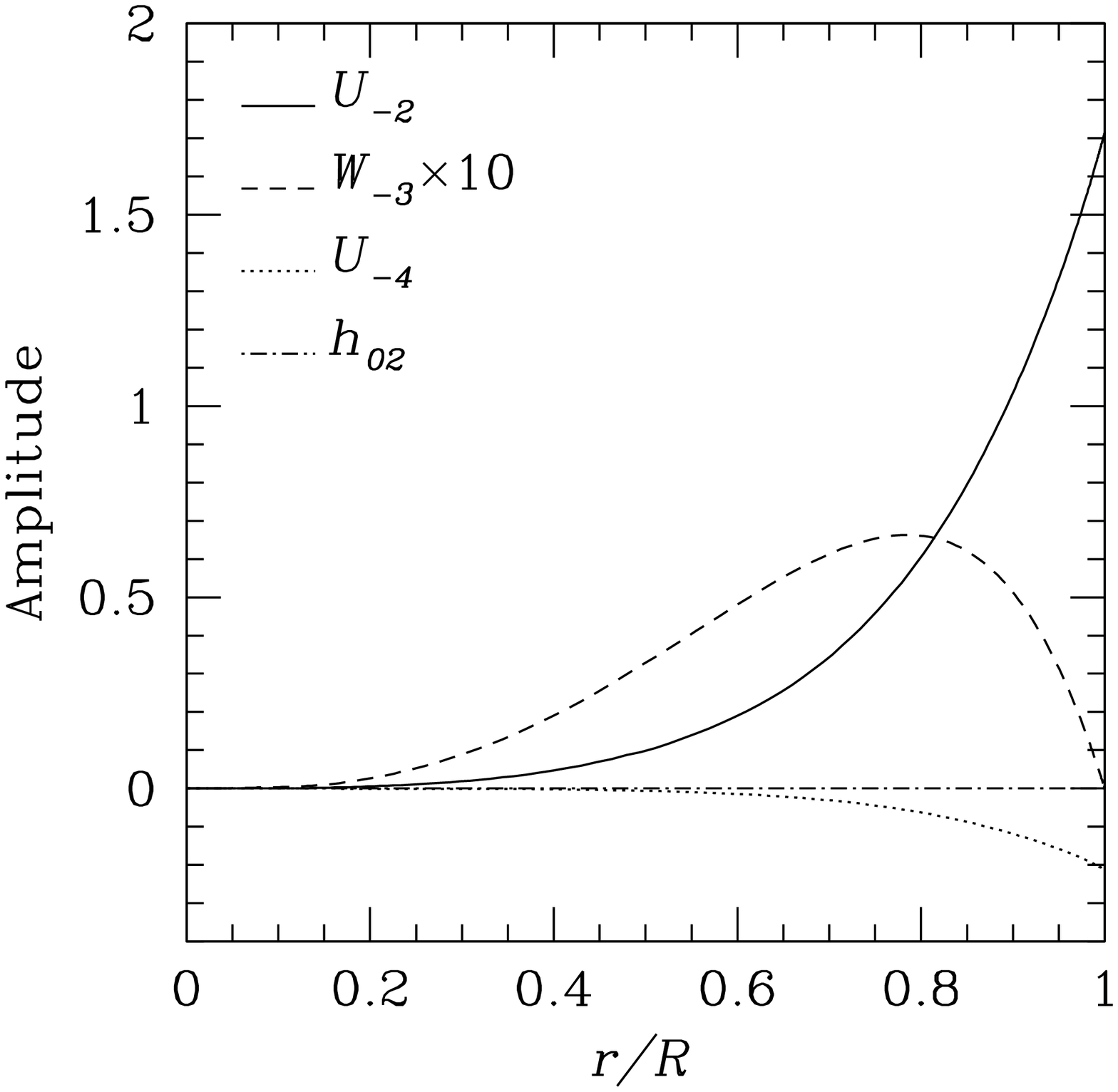}
\caption{Four dominant expansion coefficients $U_{-2}$, $W_{-3}$, $U_{-4}$,
and $h_{02}$ for the $r^s$-mode associated with $m=2$ in model II for
$\eta=0.1$, given as a function of $r/R$. Other coefficients have negligible
amplitude. The corresponding $\kappa$ is given by $\kappa=0.7240$. The 
amplitudes are normalized by $U_{n2}=1$ at $r=R$.}
\end{figure}

\end{document}